

  \magnification\magstep1
  \baselineskip = 0.5 true cm
  \parskip=0.1 true cm

  \def\sa{\vskip 0.30 true cm}
  \def\sb{\vskip 0.60 true cm}
  \def\sc{\vskip 0.15 true cm}

  \pageno = 1
  \vsize = 22.5 true cm
  \hsize = 16.1 true cm

  \font\msim=msym10
  
  \def\gc{\hbox{\msim C}}

\font\logo=logoipnl scaled \magstep4

   \line{\vbox{\hsize 4.4 true cm
   \noindent
   \logo P}\hfill \vbox{\hsize 2.8 true cm
   \noindent
   \bf LYCEN 9338\break
   November 93}}

\sb
\sa
\sb
\sc
\sb

\centerline { {\bf Point group invariants in
the $U_{qp}(u(2))$ quantum algebra picture}
\footnote{$^1$}{Dedicated to Professor R.T.~Sharp.}$^{,}$
\footnote{$^2$}{Submitted to Canadian Journal of Physics.} }

\medskip
\sa
\sb
\vskip 0.5 true cm

\centerline {M. KIBLER}

\sa

\centerline {Institut de Physique Nucl\'eaire de Lyon,}

\centerline {IN2P3-CNRS et Universit\'e Claude Bernard,}

\centerline {43 Boulevard du 11 Novembre 1918,}

\centerline {F-69622 Villeurbanne Cedex, France}

\sa
\sa
\sa
\sb
\sb

\baselineskip = 0.68 true cm

\sa


\sb

Some consequences of a $qp$-quantization of a
point group invariant developed in the enveloping
algebra of $SU(2)$ are examined in the present note.
A set of open problems concerning such invariants
in the $U_{qp}(u(2))$ quantum algebra picture
is briefly discussed.

\sc
\sa

On examine quelques unes des cons\'equences d'une
$qp$-quantification d'un invariant sous un groupe
ponctuel donn\'e d\'evelopp\'e dans l'alg\`ebre
enveloppante de $SU(2)$. On discute une s\'erie de
probl\`emes ouverts concernant de tels invariants
dans l'image de l'alg\`ebre quantique $U_{qp}(u(2))$.

\sa
\sa
\sb


\vfill\eject

\baselineskip = 0.89 true cm

\centerline {\bf 1. Introduction}

\sa

Potentials invariant under a point symmetry subgroup $G$ of
the three-dimensional rotation group $O(3)$ play an important
r\^ole in molecular physics and in the physics of crystals.
In particular, $G$-invariant operators that can be expanded in
the enveloping algebra of the two-dimensional special unitary
group $SU(2)$ are of central importance in luminescence
spectroscopy and electron paramagnetic resonance
of transition ions in crystalline environments as
well as in rotational spectroscopy of molecules and, to a less
extent, of nuclei (see for instance refs.~1-3).
The determination of operators $V_G({\cal J}_u)$
that are polynomials in the generators
${\cal J}_u = {\cal J}_x, {\cal J}_y, {\cal J}_z$
of the group $SU(2)$ and invariant under a (finite)
group $G$ may be achieved by means of the method of operator
equivalents, as first developed by Stevens (1) in the framework
of crystal-field theory (see also refs.~2-5). Furthermore,
group theoretical methods, based on the use of the so-called
Molien generating function, have been developed by several
people (6-12), involving Sharp and some of his colleagues,
for obtaining an integrity basis of operators $V_G({\cal J}_u)$.

According to Wigner theorem, the spectrum of $V_G({\cal J}_u)$
may be characterized by representation classes of $G$. More
precisely, in the absence of accidental degeneracies, the
eigenvalues $W(j, \Gamma)$ arising from the diagonalization of
$V_G({\cal J}_u)$ on a subspace
$\varepsilon (j) = \left\{ \vert jm) \ : \ m=-j, -j+1, \cdots, j \right\}$
of constant angular momentum $j$ can be, at least partially,
labelled by irreducible representation classes (IRC's) $\Gamma$
of $G$ or of its spinor group $G^*$ ($G = G^*/Z_2$).
(The vector $\vert jm)$
is a common eigenstate of the angular momentum operators
${\cal J}^2 = {\cal J}_x^2 +
              {\cal J}_y^2 +
              {\cal J}_z^2$
and ${\cal J}_z$.)

It is the aim of this short paper to examine,
from the point of view of spectral analysis,
some of the consequences of replacing the Lie
algebra $su(2)$ of the group $SU(2)$ by a
$qp$-quantized universal enveloping algebra
$U_{qp}(u(2))$. More specifically, we want to address here the
following question. What happens when we replace in
$V_G({\cal J}_u)$ each basis element ${\cal J}_u$ of $su(2)$ by
the corresponding basis element $J_u$ of the quantum algebra
$U_{qp}(u(2))$~? We shall not answer the latter question in the
general case where $V_G({\cal J}_u)$ is arbitrary. We shall
rather investigate two particular cases of deformed invariants
and shall try to
extrapolate some general tendency.

The mathematical and physical motivations for investigating
more general deformed invariants of the type
$V_G({\cal J}_u \mapsto J_u)$, in the
$qp$-quantized universal enveloping algebra $U_{qp}(u(2))$,
are strongly
connected. From a mathematical point of view, we may wonder
whether the spectrum of $V_G(J_u)$ exhibits the
degeneracies afforded by $G$. This is another way to address
the questions what is the relation between the finite group $G$
and the quantum algebra $U_{qp}(u(2))$ and what is the
relation between $G$ and the deformed invariants. We suspect
that the replacement of $V_G({\cal J}_u)$ by $V_G(J_u)$ yields
some symmetry breaking, a fact that is certainly interesting
from a physical viewpoint (we have in mind some applications to
electron paramagnetic resonance and to the Jahn-Teller effect).

\sb

\centerline {\bf 2. A two-parameter quantum algebra}

\sa

We begin with those aspects of the quantum algebra
$U_{qp}(u(2))$, recently introduced in ref.~13, that
are of relevance for section 3. Loosely speaking, the
two-parameter quantum algebra $U_{qp}(u(2))$ is spanned by the
operators $J_x$, $J_y$, $J_z$ and $J_0$ acting on the space
$\varepsilon = \bigoplus_j \varepsilon (j)$ and satisfying the
commutation relations
$$
\eqalign{
  & [J_0, J_x] = [J_0, J_y] = [J_0, J_z] = 0
\cr
  & [J_x, J_y] = {i \over 2} \> (qp)^{J_0 - J_3} \> [2J_z]_{qp}
\cr
  & [J_y, J_z] = i J_x, \quad [J_z, J_x]= i J_y
}
\eqno [1]
$$
In this paper, we use the notation
$$
[X]_{qp} = {q^X - p^X \over q-p}
\eqno [2]
$$
for both operators ($X$ acts on the Hilbert space $\varepsilon$)
and numbers ($X$ belongs to the field of complex numbers $\gc$).
By introducing $s = \ln q$ and $r = \ln p$, eq.~[2]
can be rewritten in the useful form
$$
[X]_{qp} = { \sinh (X {s-r \over 2}) \over \sinh ({s-r \over 2}) }
\> \exp \left[ (X-1) {s+r \over 2} \right]
\eqno [3]
$$
which is easy to handle when $p=q^{-1} \to 1$.

The Hopf algebraic structure of $U_{qp}(u(2))$ needs the
introduction of a coproduct, an antipode and a counit (see
ref.~13 for details). In the particular case $p=q^{-1}$
(or $r=-s$), the operators  $J_x$, $J_y$ and  $J_z$ span the
well-known quantum algebra $U_{q}(su(2))$ described by many
authors (see for example refs.~14 and 15). In addition, in the
limiting case $p=q^{-1} = 1$, the quantum algebra $U_{qp}(u(2))$
gives back the Lie algebra $u(2)$. Equation [1]
indicates that the permutational symmetry of $(x,y,z)$ is
broken when going from $u(2)$ to $U_{qp}(u(2))$.

The action on the space $\varepsilon$ of the generators
${J}_{\pm} = {J}_x \pm i {J}_y$,
${J}_3     = {J}_z$ and
${J}_0$ of $U_{qp}(u(2))$ is given by (see ref.~13)
$$
\eqalign{
  J_{\pm} \; & |jm ) \; = \; {\sqrt {[j \mp m]_{qp} \; [j \pm m + 1]_{qp}}}
\; |j, m \pm 1 )                   \cr
& J_3 \; |jm ) \; = \; m \; |jm ), \quad
  J_0 \; |jm ) \; = \; j \; |jm )  \cr
}
\eqno [4]
$$
Note that $J_+$ turns out to be the adjoint of $J_-$ if $s-r$
and $s+r$ are real numbers or if $s-r$ is a pure imaginary
number and $s+r$ a real number.

To close this section, let us mention that the operator
$$
C_2 \left( U_{qp}(u(2)) \right) \; = \;
J_x^2 + J_y^2 + {1 \over {2}} \; {[2]_{qp}} \;(qp)^{J_0 - J_z} \;
([J_z]_{qp})^2
\eqno [5]
$$
is an invariant, abbreviated as $J^2$ in the following,
for $U_{qp}(u(2))$ in the sense that it commutes
with each
of the generators $J_x$, $J_y$, $J_z$ and $J_0$. This invariant
has the eigenvalues $[j]_{qp} \> [j+1]_{qp}$.
In the limiting case
$p = q^{-1} = 1$, note that the
operator $C_2 \left( U_{qp}(u(2)) \right)$ can be identified
with the Casimir operator $C_2(su(2)) = {\cal J}^2$
of $su(2)$.

\sb

\centerline {\bf 3. Point group invariants}

\sa

We are now in a position to look at two examples for
$V_G({\cal J}_u \mapsto J_u)$. The first example is devoted to the limiting
situation where $G \equiv O(3)$. In this situation, the simplest
operator $V_{O(3)}(J_u)$ is the second-order polynomial
$$
\phi^2_{\rm axial} \> = \> J_x^2 +
                           J_y^2 +
                           J_z^2
\eqno [6]
$$
In the limiting case $p = q^{-1} = 1$, the operator
$\phi^2_{\rm axial}$ is
clearly $O(3)$-invariant since it coincides then with the
Casimir operator ${\cal J}^2$ of $su(2)$. For generic $q$ and $p$, we
suspect from eq.~[1] that $\phi^2_{\rm axial}$
is only axially invariant rather than being fully rotationally
invariant. This may be easily proved by looking for the
eigenvalues of $\phi^2_{\rm axial}$ on the subspace
$\varepsilon (j)$. We obtain
$$
W_2(j, \Gamma_{\vert m \vert}) =
{1 \over 2} \left( [j-m]_{qp} [j+m+1]_{qp}
           \> + \> [j+m]_{qp} [j-m+1]_{qp} \right) + m^2
\eqno [7]
$$
with $m = -j, -j+1, \cdots, j$. (Of course,
$W_2(j, \Gamma_{\vert m \vert}) \to j(j+1)$ when
$p=q^{-1} \to 1$.) The eigenvalues [7] are $m$-dependent
and invariant under the interchange $m \leftrightarrow -m$. The
spectrum of $\phi^2_{\rm axial}$ thus consists of $j + {1 \over 2}$ doublets
if $j$ is half of an odd integer and of one singlet and $j$
doublets if $j$ is an integer. It can be described with the
help of the IRC's
$\Gamma_{\vert m \vert}$ of
the axial group $C_{\infty v}$. (In molecular physics notations,
the IRC's of    $C_{\infty v}$ are
$\Gamma_0 = A_1$ for $j$ even, $\Gamma_0 = A_2$ for $j$ odd and
$\Gamma_{\vert m \vert} = E_{\vert m \vert}$ for $m \ne 0$.)
As a net result, the
$qp$-quantization of the operator
$V_{O(3)}({\cal J}_u) = {\cal J}^2$,
via the replacement ${\cal J}_u \mapsto J_u$,
leads to a symmetry
breaking characterized by the chain $O(3) \supset C_{\infty v}$.

The second example of operator
$V_G({\cal J}_u \mapsto J_u)$ is concerned
with the octahedral group $G \equiv O$. Let us take for
$V_O(J_u)$ the fourth-order polynomial
$$
  \phi^4_{\rm trigo} =
  {\Delta \over 180} \>
  \left\{ 5 \sqrt{2}
 [ J_z(J_+^3 + J_-^3) +
     (J_+^3 + J_-^3)J_z ]
  - 35 J_z^4 - 25 J_z^2 + 30 J_z^2J^2 + 6 J^2 - 3 J^4 \right\}
\eqno [8]
$$
where $\Delta$ is a positive parameter. (The parameter $\Delta$
has a well-known significance
in the spectroscopy of $d^N$ ions in cubical symmetry,
see below.)
In the limiting case $p=q^{-1} = 1$, the operator
$\phi^4_{\rm trigo}$ can be shown to be invariant under the
octahedral group $O$. For generic $q$ and $p$, the
diagonalization of $\phi^4_{\rm trigo}$ on a manifold
$\varepsilon (j)$ does not lead generally to a spectrum of the
type of the one afforded by the group $O$. As an illustration,
the diagonalization of $\phi^4_{\rm trigo}$ on the subspace
$\varepsilon (2)$ yields two doublets
$$
\eqalign{
  W_4 (2, \langle ET_2 \rangle E)_{+} & = {1 \over 2} \left  [
  \alpha + \beta {+} \sqrt {(\alpha - \beta)^2 + 4 \gamma ^2}
                                \right ] \cr
  W_4 (2, \langle ET_2 \rangle E)_{-} & = {1 \over 2} \left  [
  \alpha + \beta {-} \sqrt {(\alpha - \beta)^2 + 4 \gamma ^2}
                                \right ]
}
  \eqno [9]
$$
and one singlet
$$
W_4 (2, \langle T_2 \rangle A_1) = {1 \over 60} \> \Delta \> d \> (2 - d)
\eqno [10]
$$
In eqs.~[9,~10], we have put
$$
  \eqalign{
  \alpha & =
  + {1 \over 540}        \; \Delta \left( 20 \, d \, \sqrt {[4]_{qp}}
  - 780  + 198 \, d - 9 \, d^2 \right) \cr
  \beta & =
  - {1 \over 540}        \; \Delta \left( 20 \, d \, \sqrt {[4]_{qp}}
  + 1380 - 288 \, d + 9 \, d^2 \right) \cr
  \gamma & =
  + {\sqrt{2} \over 108} \; \Delta \left(       d \, \sqrt {[4]_{qp}}
  - 120  +  18 \, d            \right)
}
\eqno [11]
$$
with $d = [2]_{qp} [3]_{qp}$. The degeneracy of the spectrum
[9,~10] is characteristic of the dihedral group $D_3$, a trigonal
subgroup of the group $O$. The eigenvalues in [9,~10] are
labelled by the IRC's $A_1$ and $E$ of the group $D_3$. We have
also indicated in [9,~10],
within $\langle \ \rangle$,
the parent IRC's $T_2$ and $E,T_2$ of
the group $O$ for the trigonal levels of symmetry $A_1$ and
$E$, respectively. In the limiting case $p=q^{-1} = 1$, we get
$$
  W_4 (2, \langle ET_2 \rangle E  )_{+}  = + {3 \over 5} \Delta,
\quad
  W_4 (2, \langle ET_2 \rangle E  )_{-}  = - {2 \over 5} \Delta,
\quad
  W_4 (2, \langle T_2  \rangle A_1)      = - {2 \over 5} \Delta
\eqno [12]
$$
The trigonal levels [9,~10] thus reduce to
the cubical levels $W_4(2, E ) =+(3/5) \Delta$ and
                   $W_4(2,T_2) =-(2/5) \Delta$ of cubical symmetry $E$ and
$T_2$, respectively. (Observe that
$\Delta = W_4(2, E ) -
          W_4(2,T_2)$.) In other words, passing from the
limiting case where $p=q^{-1} =1$ to the case where $q$ and $p$
are arbitrary, produces a level splitting
$$
    E   \rightarrow E, \qquad
    T_2 \rightarrow A_1 \oplus E
\eqno [13]
$$
corresponding to the chain of groups $O \supset D_3$.

\sb

\centerline {\bf 4. Concluding remarks}

\sa

Five concluding remarks may be drawn from the results of
sections 2 and 3.

(i) As a trivial remark, we note that the Casimir invariant
$C_2(su(2)) = {\cal J}^2$ of the Lie algebra $su(2)$ is
also an invariant of the quantum algebra $U_{qp}(u(2))$~; this
may be easily checked from the commutation relations of
${\cal J}^2$ with ${J}_u$ ($u = x, y, z, 0$). However, a
$qp$-quantization of ${\cal J}^2$ leads to an operator, the
operator $\phi^2_{\rm axial}$ of eq.~[6], that is not an
invariant of $U_{qp}(u(2))$. Such a $qp$-quantization yields a
symmetry breaking described by the restriction of $O(3)$ to its
subgroup $C_{\infty v}$.

(ii) An incomplete reciprocal part of point
(i) is as follows~: the invariant operator
$C_2(U_{qp}(u(2))) = J^2$ of the quantum algebra $U_{qp}(u(2))$ is
also an invariant operator of the Lie algebra $su(2)$.

(iii) In complement of points (i) and (ii), it is
possible to show that the invariant $J^2$ of $U_{qp}(u(2))$
can be expressed by series involving the invariants ${\cal J}^2$ and
${\cal J}_0$ of the Lie algebra $u(2)$ (see ref.~16). For instance, in
the particular case where $p=q^{-1}$ with $q = \exp (i \varphi) \in S^1$,
we can prove that
$$
J^2 = \sum_{k=1}^{+\infty} \> a_k(\varphi) \> \left( {\cal J}^2 \right)^k
\eqno [14]
$$
where
$$
a_k(\varphi) = 2^{2 k -1 }   \> {1 \over \sin^2 \varphi} \>
 \sum_{\ell =0}^{+\infty }   \>
      (-1)^{k + \ell + 1 }   \>
  \varphi^{ 2(k + \ell)  }   \>
{ 1 \over {[2(k + \ell)]!} } \>
{ {(k + \ell)!} \over {k! \> \ell!} }
\eqno [15]
$$
In terms of the spherical Bessel functions of
the first kind $j_{k-1}$, we have
$$
a_k(\varphi) = (-1)^{k - 1 } \> { \varphi^2 \over {\sin^2 \varphi} } \>
 { (2 \varphi)^{ k - 1 } \over k! } \> j_{k-1}(\varphi)
\eqno [16]
$$
The transcription of eqs.~[14,~16] in terms of eigenvalues gives
the formula derived in ref.~17 in connection with rotational
spectroscopy of nuclei. Note that the generalization of [14-16]
to any
doublet $(U_{qp}(g),g)$, $g$ being a simple Lie algebra, is an
open problem.

(iv) The $qp$-quantization of a $G$-invariant operator $V_G({\cal J}_u)$,
through the replacement ${\cal J}_u \mapsto J_u$, produces an
operator that is invariant under a subgroup $H$ of $G$ rather
than being invariant under $G$. A question then naturally
arises: given a group $G$, what $G \supset H$ symmetry breaking
do we obtain by performing a $qp$-quantization of an operator
of type $V_G({\cal J}_u)$ (see also ref.~18)~? The answer
is certainly not unique. Let us clarify the latter assertion.
In the case of $\phi^4_{\rm trigo}$, we have obtained an
$O \supset D_3$ symmetry breaking corresponding to
$H \equiv  D_3$. This comes from the fact that the operator
$V_O({\cal J}_u)$ implicitly considered in section 3 is a
cubical invariant oriented according to a $C_3$ axis. Should we
have considered an operator $W_O({\cal J}_u)$ oriented
according to a $C_4$ principal axis, equivalent to
$V_O({\cal J}_u)$ as far as their spectra are concerned, we
would have obtained an $O \supset D_4$ symmetry breaking
corresponding to the tetragonal subgroup $H \equiv D_4$ of $O$
(cf.~ref.~18). In this respect, it should be interesting to
investigate the $G \supset H$ symmetry breakings that we may
obtain from the $qp$-quantization of the integrity basis for
the operators $V_G({\cal J}_u)$ derived in refs.~6-12.

(v) Finally, to find in a systematic way polynomials, in the
generators $J_u$ ($u = x,y,z,0$) of the quantum algebra
$U_{qp}(u(2))$, that are invariant under a finite subgroup $G$
of $O(3)$ is an appealing problem. The generating function
methods, applied in refs.~6-12 to the derivation of
operators $V_G({\cal J}_u)$, might be a key for solving this
problem.

Points  (i)-(v) pave the way for future investigations and
we hope to return on these matters in a forthcoming paper.

To close this paper, it should be mentioned that multiparameter
quantum algebras have been studied by many authors including
Sudbery (19) and Fairlie and Zachos (20) among others
(see also references
in refs.~13 and 16). The main advantage of our version of the
quantum algebra
 $U_{qp}(u(2))$
introduced in section 2
lies in the fact that the latter algebra is not equivalent to a
one-parameter algebra as far as the eigenvalues of the Casimir
operator [5] are concerned. The reader should consult ref.~16
where this fact has been successfully used to develop a model
for rotational spectroscopy of nuclei.

\sb

\centerline {\bf Acknowledgement}

\sa

The present paper is dedicated to Bob Sharp. The author very much
enjoyed the nice lectures by Professor R.T.~Sharp when he was a
post-doctoral fellow in the {\it Department of Mathematics of
McGill University} in 1969-1970. Since 1976, the
author benefited from discussions and seminars with Bob Sharp
on the occasion of several stays at the {\it Centre de
recherches math\'ematiques de l'Universit\'e de
Montr\'eal}. Merci Bob.

\vfill\eject

\centerline {\bf References}

\sa
\sa

\baselineskip = 0.60 true cm

\noindent
\item{1.} K.W.H.~Stevens,
Proc. Phys. Soc. London, Sect A, {\bf 65}, 209 (1952).

\noindent
\item{2.} M.R.~Kibler.
{\it In} Group theoretical methods in physics.
{\it Edited by} R.T. Sharp and B. Kolman,
Academic Press, Inc., New York, NY. 1977. (p.~161-172)

\noindent
\item{3.} M. Kibler and R. Chatterjee,
Can. J. Phys. {\bf 56}, 1218 (1978).

\noindent
\item{4.} M.R.~Kibler.
{\it In} Recent advances in group theory and their
application to spectroscopy. {\it Edited by} J.C.
Donini, Plenum Press, New York, NY. 1979. (p.~1-96)

\noindent
\item{5.} M. Kibler and G. Grenet,
J. Math. Phys. {\bf 21}, 422 (1980).

\noindent
\item{6.} J.~Patera and P.~Winternitz, J. Chem. Phys.
{\bf 65}, 2725 (1976).

\noindent
\item{7.} L.~Michel.
{\it In} Group theoretical methods in physics.
{\it Edited by} R.T. Sharp and B. Kolman,
Academic Press, Inc., New York, NY. 1977. (p.~75-91)

\noindent
\item{8.} P.~Winternitz.
{\it In} Group theoretical methods in physics.
{\it Edited by} R.T. Sharp and B. Kolman,
Academic Press, Inc., New York, NY. 1977. (p.~549-572)

\noindent
\item{9.} R.~Gaskell, A.~Peccia and R.T.~Sharp,
J. Math. Phys. {\bf 19},  727 (1978).

\noindent
\item{10.} J.~Patera, R.T.~Sharp and P.~Winternitz,
J. Math. Phys. {\bf 19}, 2362 (1978).

\noindent
\item{11.} P.E.~Desmier and R.T.~Sharp, J. Math. Phys.
{\bf 20}, 74 (1979).

\noindent
\item{12.} J.~Patera and R.T.~Sharp.
{\it In} Recent advances in group theory and their
application to spectroscopy. {\it Edited by} J.C.
Donini, Plenum Press, New York, NY. 1979. (p.~219-248)

\noindent
\item{13.} M.R. Kibler.
{\it In} Symmetry and structural properties of
condensed matter. {\it Edited by} W. Florek,
D. Lipi\'nski and T. Lulek,
World Scientific, Singapore. 1993. (p.~445-464)

\noindent
\item{14.} L.C.~Biedenharn, J. Phys. A, {\bf 22}, L873 (1989).

\noindent
\item{15.} A.J.~Macfarlane, J. Phys. A, {\bf 22}, 4581 (1989).

\noindent
\item{16.} R.~Barbier, J.~Meyer and M.~Kibler,
submitted for publication.

\noindent
\item{17.} D. Bonatsos, E.N. Argyres, S.B. Drenska,
P.P. Raychev, R.P. Roussev and Yu.F. Smirnov,
Phys. Lett. B, {\bf 251}, 477 (1990).

\noindent
\item{18.} M.~Kibler and J.~Sztucki, Acta Phys. Pol. A, in press.

\noindent
\item{19.} A.~Sudbery, J. Phys. A, {\bf 23}, L697 (1990).

\noindent
\item{20.} D.B.~Fairlie and C.K.~Zachos, Phys.~Lett.~B, {\bf 256}, 43 (1991).

\bye